\documentclass[aps,prl,twocolumn,superscriptaddress,amsmath,amssymb,amsfonts,10pt]{revtex4-2}

\usepackage{graphicx}
\usepackage{braket}
\usepackage[dvipsnames]{xcolor}
\usepackage[normalem]{ulem}
\usepackage[bookmarks=false,colorlinks=true,urlcolor=RoyalBlue,citecolor=NavyBlue,linkcolor=OrangeRed]{hyperref}
\usepackage{bm}

\begin{document}

\title{Symmetry Fragmentation}

\author{Thomas Iadecola}
\email{iadecola@psu.edu}
\affiliation{Department of Physics and Astronomy, Iowa State University, Ames, Iowa 50011, USA}
\affiliation{Ames National Laboratory, Ames, Iowa 50011, USA}
\affiliation{Department of Physics, The Pennsylvania State University, University Park, PA 16802, USA}
\affiliation{Institute for Computational and Data Sciences, The Pennsylvania State University, University Park, PA 16802, USA}
\affiliation{Materials Research Institute, The Pennsylvania State University, University Park, PA 16802, USA}

\date{\today}

\begin{abstract}
In quantum many-body systems with kinetically constrained dynamics, the Hilbert space can split into exponentially many disconnected subsectors, a phenomenon known as Hilbert-space fragmentation.
We study the interplay of such fragmentation with symmetries, focusing on charge conserving systems with charge conjugation and translation symmetries as a concrete example.
The non-Abelian algebra of these symmetries and the projectors onto the fragmented subsectors leads to the emergence of exponentially many logical qubits encoded in degenerate pairs of eigenstates, which can be highly entangled.
This algebra also provides necessary conditions for experimental signatures of Hilbert-space fragmentation, such as the persistence of density imbalances at late times.
\end{abstract}

\maketitle

In an isolated quantum many-body system, a short-range-correlated initial state generically evolves toward a late-time state whose local properties can be predicted by an appropriate statistical ensemble~\cite{Deutsch91,Srednicki94,Rigol08,D'Alessio16,Deutsch18}. 
When this ensemble derives from standard statistical mechanics, only the large-scale features of the initial state (e.g., its energy, total charge, etc.) are relevant, while the details (e.g., the locations of the charges) are lost.
Quantum superpositions also lose their coherence under such dynamics, since the superposition components generically project onto Hamiltonian eigenstates with many different energies.
Thus, quantum dynamics tends to erase both classical and quantum information present in the initial state as the system thermalizes.

Many mechanisms have been proposed whereby quantum systems can retain initial-state information.
For example, strongly disordered systems can preserve classical information about the initial locations of the particles~\cite{Anderson58,Basko06,Oganesyan07,Pal10,Huse14,Serbyn13a,Serbyn13b,Nandkishore15,Abanin19}.
Systems with strong zero modes~\cite{Fendley12,Bahri15,Fendley16,Else17a,Kemp17,Kemp20,Jin25} can preserve quantum information due to the presence of one or more \textit{encoded qubits}, which guarantee a pattern of degeneracies among the eigenstates that prevents the decoherence of certain initial superpositions.
This information can be spatially localized on the boundary of a quantum system with a thermalizing bulk~\cite{Else17a,Kemp17,Rakovszky20,Kemp20,Jin25}, or, in the presence of some additional structure defining a subsystem code~\cite{Bacon2005a,Poulin2005}, it can be delocalized across an otherwise thermalizing system~\cite{Wildeboer22}.
There is a limit to the amount of quantum information stored in this way---the number of encoded qubits is typically constant in the number of physical qubits participating in the dynamics.

\begin{figure}[b!]
    \centering
    \includegraphics[width=\columnwidth]{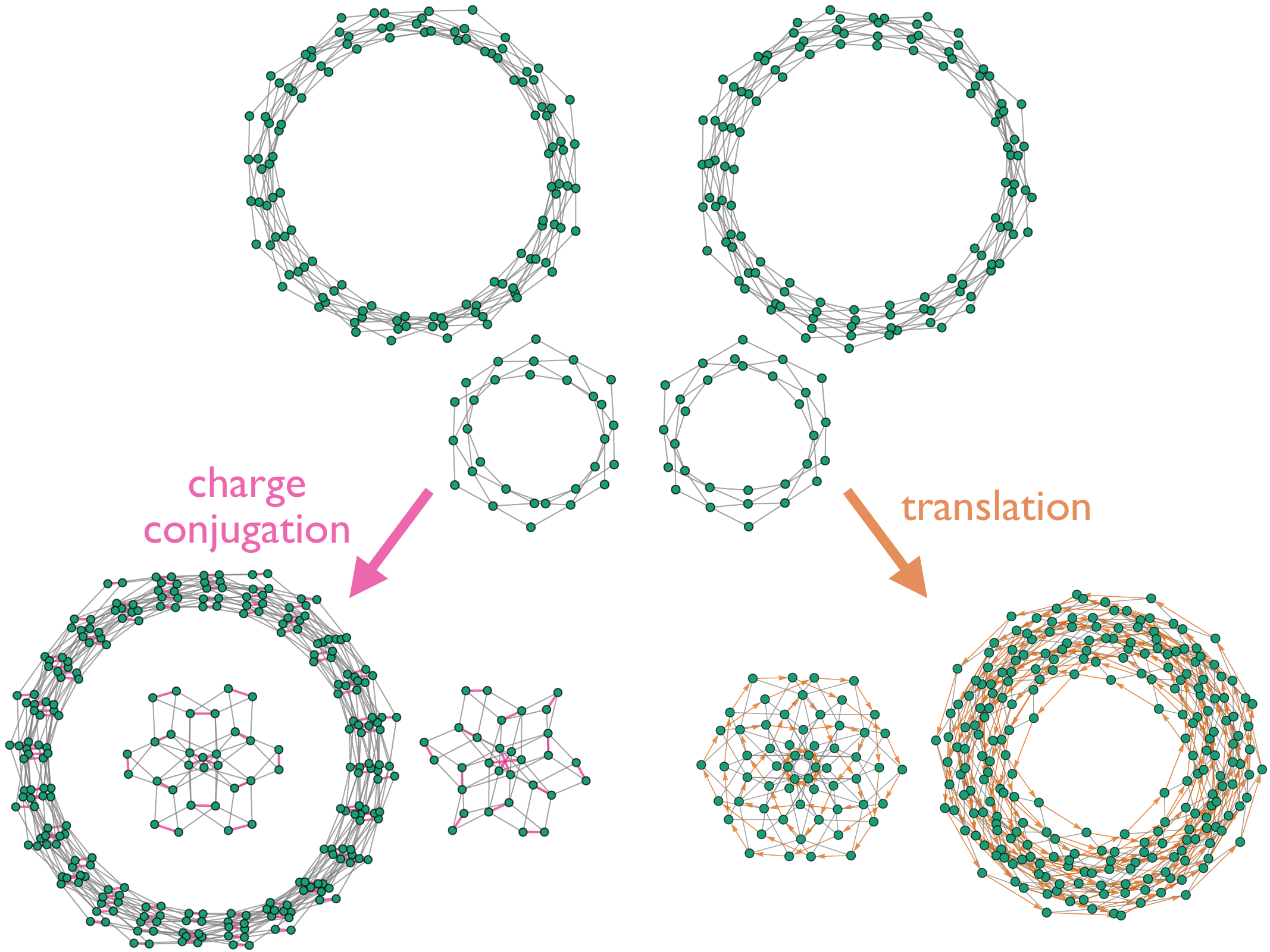}
    \caption{
    Symmetry fragmentation. Top: Interaction graph of the Hamiltonian \eqref{eq:xnor} showing four Krylov sectors. Charge conjugation symmetry (pink bonds) pairs the two largest Krylov sectors together while leaving the smaller two invariant. Translation symmetry (orange arrows) collapses both pairs of sectors, since neither is translation invariant.
    }
    \label{fig:graph}
\end{figure}

In this paper, we describe a mechanism whereby an \textit{exponential} number of encoded qubits can be stored in a single many-body system, coexisting with a restricted notion of thermalization. 
The mechanism relies on the phenomenon of Hilbert-space fragmentation (HSF)~\cite{Sala20,Khemani20,Moudgalya22a,Chandran23}, in which the Hilbert space of a many-body system breaks into exponentially many dynamically disconnected subspaces sometimes called Krylov sectors.
HSF is ubiquitous in systems with kinetic constraints~\cite{vanHorssen15,Hickey16,Lan18,Pancotti20,Brighi23,Ganguli25,Zhao25}, which can be intrinsic or induced by symmetries~\cite{Sala20,Khemani20,Rakovszky20,Moudgalya21a} or extrinsic potentials~\cite{Khemani20,Scherg21,Desaules21,Kohlert23}, and in strongly-confined gauge theories~\cite{Yang20,Chen21,Desaules24,Desaules25}.
In the simplest cases, which are called classically fragmented~\cite{Moudgalya22b}, the Krylov subspaces consist of product states reachable from a given initial computational basis state; the subspace to which the initial state belongs is then a form of classical information that is preserved under the dynamics.
More complicated ``quantum HSF" can occur in non-product-state bases~\cite{Moudgalya22b,Brighi23,Ganguli25} but does not generically guarantee the existence of encoded qubits.

This work demonstrates that exponentially many encoded qubits can emerge from the interplay of HSF and conventional symmetries. 
Such symmetries need not leave each Krylov sector invariant, and may instead map pairs of sectors onto each other (see Fig.~\ref{fig:graph}) or mix them in more complicated ways.
Focusing on U(1) charge-conserving systems with a $\mathbb Z_2$ charge conjugation symmetry, we demonstrate that the nontrivial action of the $\mathbb Z_2$ symmetry on Krylov sectors leads to exponentially many twofold degeneracies by way of an emergent SU(2) algebra.
These degeneracies correspond to the formation of simple Schr\"odinger cat-like eigenstates with long-range correlations, and more complex eigenstates that are superpositions of highly entangled eigenstates of individual Krylov sectors.
The presence of these superpositions can be probed in quantum dynamics experiments.
Further interplay of this structure with translation symmetry leads to necessary criteria for certain experimental signatures of HSF, including the presence of particle density imbalances at late times.

\textit{Motivating example.}---Consider the one-dimensional quantum XY model 
\begin{align}
    H_{\rm XY}=\frac{1}{2}\sum^L_{i=1}( X_iX_{i+1}+Y_iY_{i+1}),
\end{align}
where $X_i,Y_i,Z_i$ are Pauli operators on site $i$, and where we assume periodic boundary conditions (PBC) $L+1\equiv 1$ without loss of generality. 
To exemplify the mechanism at play, we consider the U(1)$\rtimes \mathbb Z_2$ symmetry group generated by the magnetization (i.e., total charge) operator $M=\sum^L_{i=1} Z_i$ and the operator $X=\prod^L_{i=1}X_i$ (i.e., charge conjugation).
These symmetry generators do not commute---$X$ maps the subspace with magnetization $m$ onto the one with magnetization $-m$.
Thus, if we choose to diagonalize $H$ in the eigenbasis of $M$ (i.e., the computational basis), then we are forced to break charge conjugation symmetry unless we work in the sector with $m=0$.
Conversely, to diagonalize $H$ in the eigenbasis of $X$ we can form Schr\"odinger cat states $(\ket{b}\pm X\ket{b})/\sqrt{2}$, where $\ket{b}$ is a product state in the magnetization-$m$ sector and $X\ket{b}$ is its charge conjugate.
The ability to diagonalize $H$ in either basis imposes a strong constraint on the spectrum: the sectors with magnetization $\pm m$ must have identical energy eigenvalues. 
In other words, each energy eigenstate $\ket{E,m}$ in the magnetization-$m$ sector must have a partner state $\ket{E,-m}$ with the same energy $E$.
As an extreme example, pairing the sectors with magnetization $m=\pm L$ results in the Greenberger-Horne-Zeilinger (GHZ) state $(\ket{00\dots}+\ket{11\dots})/\sqrt{2}$, a paradigmatic example of many-body entanglement and long-range correlations, as a mid-spectrum eigenstate of $H_{\rm XY}$ with $E=0$.

These degeneracies derive from an emergent SU(2) algebra.
Let $P$ and $Q$ denote the orthogonal projectors onto the magnetization-$m$ and $-m$ sectors, respectively.
Then, we can define operators
\begin{align}
\begin{split}
\label{eq:logical}
    \mathcal I &= P+Q,\indent \mathcal Z = P - Q,\\
    \mathcal X &= XP + XQ,\indent i\mathcal Y = XQ-XP
\end{split}
\end{align}
that obey
\begin{align}
\begin{split}
\label{eq:algebra}
\mathcal X^2 &= \mathcal Y^2 = \mathcal Z^2 = \mathcal I\\    
\{\mathcal X,\mathcal Y\}&=\{\mathcal X,\mathcal Z\}=\{\mathcal Y,\mathcal Z\}=0\\
[\mathcal X,\mathcal Y]&=2i\, \mathcal Z \indent (\text{+ cyclic permutations})
\end{split}
\end{align} 
as can be shown using $XPX=Q$ and the fact that $P,Q$ are orthogonal projection operators.
Given a Hamiltonian eigenstate $\ket{E,m}$, for which $\mathcal Z \ket{E,m}=\ket{E,m}$, this algebra implies the existence of the degenerate partner $\ket{E,-m}=\mathcal X\ket{E,m}$ having $\mathcal Z$ eigenvalue $-1$.

We close this example with two comments. First, the spectral degeneracy enforced by the algebra \eqref{eq:algebra} can be viewed as encoding a ``logical qubit" with corresponding generators \eqref{eq:logical}.
In a chain of (even) length $L$, there are $L/2$ such encoded qubits associated with each pair of nonzero magnetization eigenvalues $\pm m$.
[Note that the generators \eqref{eq:logical} associated with different values of $|m|$ commute.]
However, logical superposition states like $(\ket{E,m}\pm \ket{E,-m})/\sqrt{2}$ may not be easy to access as they are generically superpositions of states with macroscopically different magnetizations.
Meanwhile the $m=0$ sector, where $M$ and $X$ can be simultaneously diagonalized, does not host such encoded qubits.
Second, we note that the logical generators $\mathcal X_m$ associated with the magnetization sectors $\pm m\neq 0$ obey $X\Gamma+\sum^L_{m=1} \mathcal X_m=X$, where $\Gamma$ is the projector onto the $m=0$ sector. 
Thus, in the presence of both U(1) and $\mathbb Z_2$ symmetries, the $\mathbb Z_2$ symmetry generator breaks into pieces associated with the different (pairs of) magnetization sectors, each of which commutes independently with $H$.

\textit{Adding kinetic constraints.}---The same algebraic structure emerges with more striking consequences in systems with Hilbert space fragmentation.
For expository purposes we focus here on the so-called XNOR model~\cite{Yang20,Zadnik21a,Zadnik21b,Pozsgay21,Singh21},
\begin{align}
\label{eq:xnor}
    H_{\rm XNOR} = \frac{1}{2}\sum^L_{i=1} P^{\rm XNOR}_{i-1,i+2}\,  (X_iX_{i+1}+Y_iY_{i+1}),
\end{align}
where the projection operator $P^{\rm XNOR}_{i-1,i+2}=(1+Z_{i-1}Z_{i+2})/2$ implements a kinetic constraint whereby an XY exchange is performed between sites $i,i+1$ only if the outer neighbors on sites $i-1$ and $i+2$ have even parity.
This model exhibits HSF within each magnetization sector induced by the combination of magnetization conservation and the conservation of the number of Ising domain walls, $N_{\rm DW} = \sum_{i}(1-Z_{i}Z_{i+1})/2$, which is enforced by the local XNOR constraint.
It arises as an effective description in the strongly confining limit of the one-dimensional $\mathbb Z_2$ lattice gauge theory~\cite{Yang20}, which can in turn be realized by Rydberg atoms in the so-called antiblockade regime~\cite{Ostmann19}. 
While the XNOR model itself is integrable~\cite{Zadnik21a,Zadnik21b,Pozsgay21}, it becomes nonintegrable in the presence of a second-neighbor interaction $\Delta\sum_i Z_{i}Z_{i+2}$ that does not affect the fragmentation structure~\cite{Yang20}.

The XNOR model exhibits the same non-Abelian interplay between $M$ and $X$ as in the XY model.
(The conservation of $N_{\rm DW}$ is irrelevant for this discussion since $[N_{\rm DW},X]=0$---we may fix an eigenvalue of $N_{\rm DW}$ without loss of generality.)
However, new features emerge in the $m=0$ sector where charge conjugation previously had a trivial action.
HSF brings about extra conserved quantities, namely the projection operators onto the different Krylov sectors in a fixed $(M,N_{\rm DW})$ eigenspace.
These Krylov sectors, like the magnetization sectors of the XY model, can either transform trivially under $X$ (i.e., map onto themselves) or be exchanged by it (see Fig.~\ref{fig:graph} for an example).
Let $K=K_0+2K_\pi$ denote the number of Krylov sectors in a fixed $(M,N_{\rm DW})$ eigenspace, with $K_0$ the number of sectors that are invariant under $X$ and $K_\pi$ the number of pairs of sectors exchanged by $X$.
Then, we can define projectors $\{\Gamma_j\}^{K_0}_{j=1}$ onto the invariant sectors and $\{P_j,Q_j\}^{K_\pi}_{j=1}$ onto the paired sectors.
This gives rise to $K_\pi$ independent copies of the algebra \eqref{eq:algebra} with generators $\{\mathcal I_j,\mathcal X_j,\mathcal Y_j,\mathcal Z_j\}^{K_\pi}_{j=1}$ directly analogous to those in Eq.~\eqref{eq:logical}.
Each copy of this algebra enforces an exact degeneracy between the energy spectra of the paired Krylov sectors, corresponding to an encoded qubit associated to that pair of sectors.
These qubits are stable to any perturbations that preserve the HSF structure as well as the $\mathbb Z_2$ symmetry (e.g., any diagonal interaction term with an even number of Pauli-$Z$ operators).

The number of such encoded qubits grows exponentially with system size.
In particular, within the $m=0$ sector there are exponentially many ``frozen states," i.e., computational basis states that are annihilated by $H_{\rm XNOR}$.
These frozen states must transform nontrivially under charge conjugation, since each of them is a one-dimensional Krylov sector; thus, the number of paired sectors $2K_\pi$ is at least as large as the number of frozen states.
In addition to the two N\'eel states $\ket{0101\dots}$ and $\ket{1010\dots}$, the frozen states consist of computational basis states without nearest-neighbor domain walls~\cite{Yang20}, corresponding to the set of PBC bit strings of length $L$ with equal numbers of $0$s and $1$s such that the motifs $010$ or $101$ do not appear. 
The number of such bit strings grows asymptotically as $(\varphi+1)^L$, where $\varphi$ is the golden ratio~\cite{Munarini03}, providing an exponentially large lower bound on the number of paired Krylov sectors.

The HSF structure imprints itself on the symmetry generator $X$ via the relation
\begin{align}
    \sum^{K_0}_{j=1}X\Gamma_j + \sum^{K_\pi}_{j=1}\mathcal X_j + \sum^L_{m=1}\mathcal X_m = X.
\end{align}
Here, the last term represents the pairing of the magnetization sectors with $m\neq 0$ discussed in the previous section, and the first two terms represent the fragmentation of the $m=0$ sector into Krylov sectors.
Each of the exponentially many ``symmetry fragments" $X\Gamma_j$ and $\mathcal X_j=XP_j+XQ_j$ independently commute with $H$ and with each other, reflecting the additional structure in the $m=0$ sector imposed by the kinetic constraint.

The symmetry fragmentation described above provides a simple mechanism for long-range correlations in eigenstates.
For a one-dimensional Krylov sector containing a frozen state $\ket{\Phi}$, the charge-conjugation-symmetric eigenstates are Schr\"odinger cat states of the form $(\ket{\Phi}+X\ket{\Phi})/\sqrt{2}$, which exhibit GHZ-like long-range correlations.
More elaborately, we can consider generic Krylov sectors, which in the presence of a nonzero second-neighbor interaction $\Delta$ are nonintegrable, and typical eigenstates within such a sector are volume-law entangled~\cite{Yang20}.
Any such eigenstate within a Krylov sector with projector $P_j$, which we denote $\ket{E,+j}$, has a degenerate partner $\ket{E,-j}=\mathcal X_j\ket{E,+j}$ in the sector with projector $Q_j$, from which a cat state with volume-law entangled components can be formed.
Such states have been noted previously to arise from strong zero modes~\cite{Else17a,Else20} or an algebraic structure defining a subsystem code~\cite{Wildeboer22}, but here they find a new realization due to the interplay of HSF and symmetry.

\begin{figure}[t!]
    \centering
    \includegraphics[width=\columnwidth]{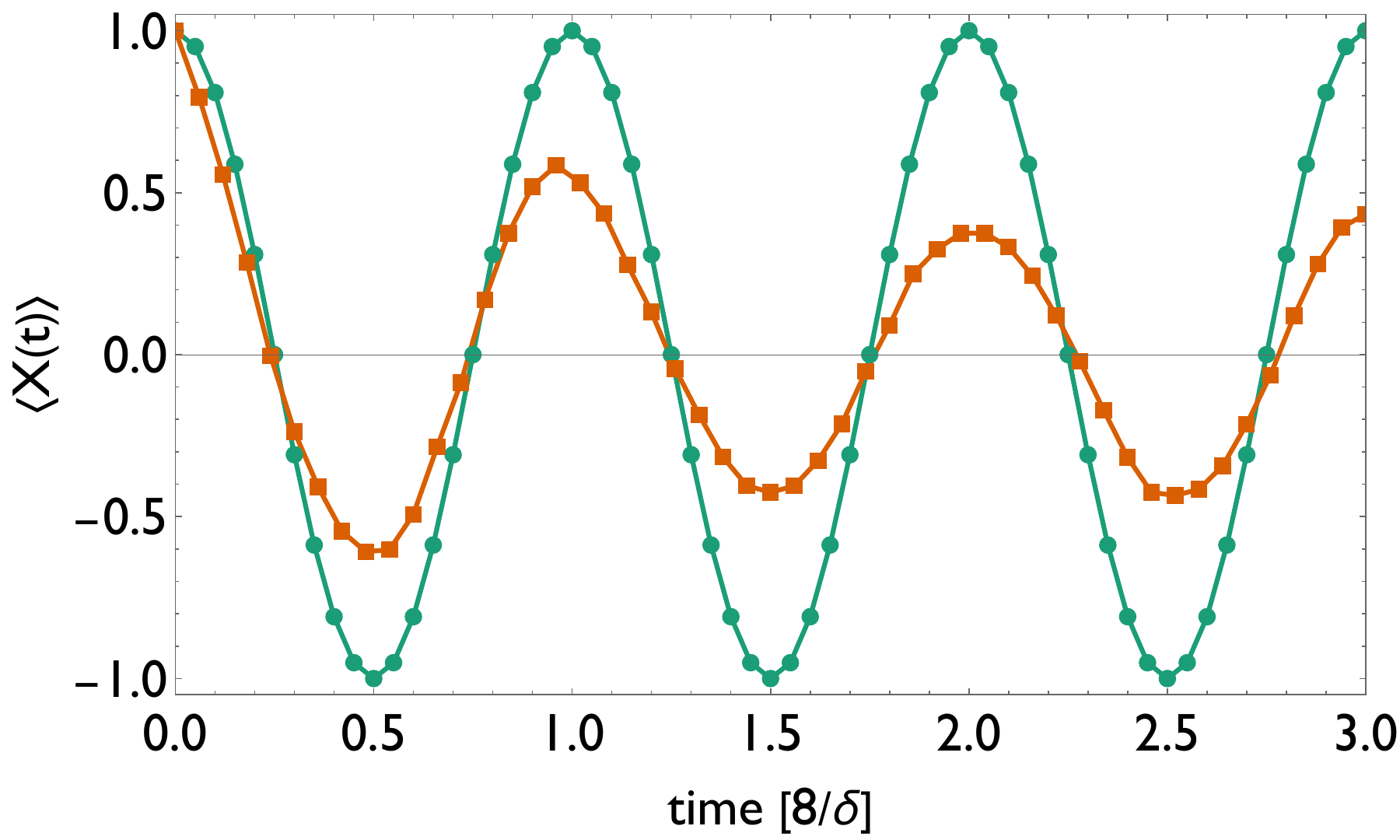}
    \caption{
Dynamics of encoded qubits under perturbed evolution at $L=18$. The time-evolved charge conjugation operator $\braket{X(t)}$ (see text) exhibits oscillations with period $8/\delta$, where $\delta$ is the qubit splitting induced by the perturbation. An initial state formed by superposing two frozen states of the unperturbed dynamics (green circles) exhibits coherent oscillations under the perturbed dynamics, while an initial superposition of two complex eigenstates (orange squares) exhibits damped oscillations.
    }
    \label{fig:qubit-dynamics}
\end{figure}

\textit{Dynamical signatures.}---The spectral degeneracy described above can be probed directly in quantum dynamics experiments via eigenstate preparation. 
For example, the GHZ-type states formed by superposing frozen states can be prepared in logarithmic depth with a unitary circuit~\cite{Cruz19} or in constant depth using midcircuit measurements~\cite{Buhrman24,Baumer24} on a gate-based quantum computer, which can then evolve the state using a Trotter circuit to demonstrate its stationarity. 
Lifting the degeneracy by adding a perturbation that breaks charge conjugation symmetry while preserving the HSF structure induces coherent many-body Rabi oscillations that can also be measured by tracking the evolution of the symmetry generator $X$ under the perturbed evolution.
As an example, Fig.~\ref{fig:qubit-dynamics} shows dynamics generated by the evolution operator $U(t)=e^{-i(\pi/4)Ht}$, with $H=H_{\rm XNOR}+\Delta\sum_{i}Z_iZ_{i+2}+h\sum_i (-1)^iZ_i$.
We set $\Delta=0.25$ to break integrability and the modulated field strength $h=0.1$ to break charge conjugation symmetry while preserving the HSF structure and plot $\braket{X(t)}=\braket{\psi_0|U^\dagger(t)\, X\, U(t)|\psi_0}$ (where $t$ is an integer) for system size $L=18$.
The initial state $\ket{\psi_0}=(\ket{\Phi}+X\ket{\Phi})/\sqrt{2}$ where $\ket{\Phi}=\ket{110001100111001100}$ is a frozen state.
The green circles show coherent oscillations with frequency $\delta=2|h\braket{\Phi|\sum_i(-1)^iZ_i|\Phi}|$ set by the perturbation-induced splitting.

For large Krylov sectors forming $\mathbb Z_2$-conjugate pairs, one can imagine preparing an eigenstate of the form $(\ket{E,+j}+\ket{E,-j})/\sqrt{2}$ and then evolving under the perturbed Hamiltonian before measuring $X$.
In this case, the initial state couples to many other eigenstates in the same Krylov sector, leading to dephasing.
An example of this is shown in Fig.~\ref{fig:qubit-dynamics}, where orange squares represent $\braket{X(t)}$ for an initial state $\ket{\psi_0}$ derived from a mid-spectrum eigenstate in a Krylov sector with 2970 states at $L=18$.
While this contrast is interesting, initial state preparation is prohibitive in this case, since it entails the creation of a coherent superposition of two eigenstates that are generically volume-law entangled.
Nevertheless, when $\mathbb Z_2$ symmetry is enforced, more experimentally accessible initial states like $(\ket{b}+X\ket{b})/\sqrt{2}$ (where $\ket{b}$ is a product state in the image of $P_j$) are superpositions of degenerate cat eigenstates $(\ket{E,+j}+\ket{E,-j})/\sqrt{2}$, and therefore retain coherence in the energy eigenbasis to arbitrarily late times.

\begin{figure}[t!]
    \centering
    \includegraphics[width=\columnwidth]{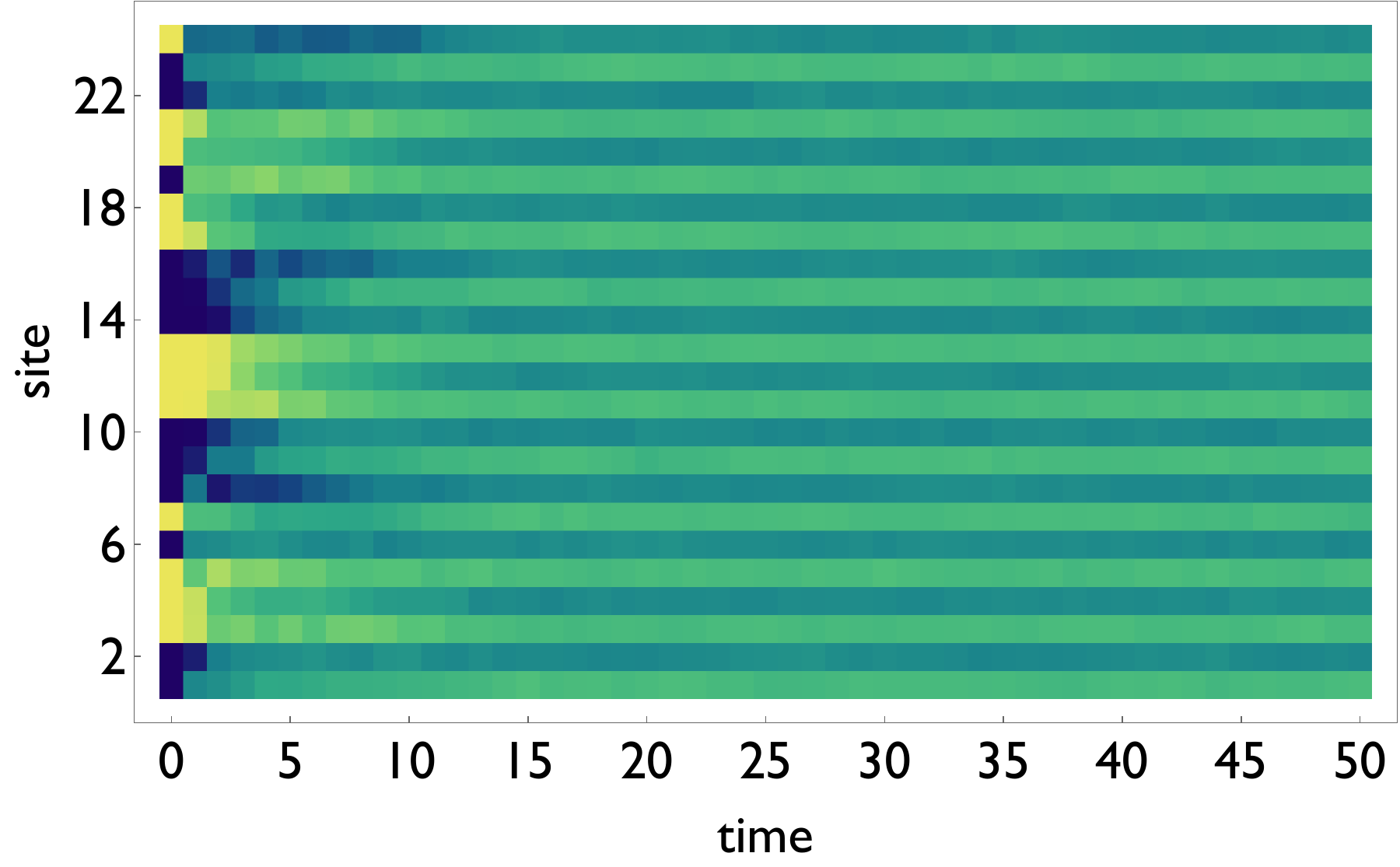}
    \caption{
    Site-resolved dynamics of an initial computational basis state on $L=24$ sites. The local expectation value $\braket{Z_i(t)}$ is represented by a color scale with light (dark) corresponding to the value $+1$ ($-1$). The initial state belongs to a Krylov sector of dimension 1456 that is neither charge-conjugation nor translation invariant, resulting in relaxation to a staggered magnetization profile at late times.
    }
    \label{fig:dynamics}
\end{figure}

Additionally, symmetry fragmentation provides necessary conditions for one of the most common experimental signatures of HSF, namely the persistence of density imbalances at late times~\cite{Guardado-Sanchez21,Kohlert23,Adler24,Wang25,Honda25,Zhao25}.
This results from the interplay of HSF with charge conjugation and spatial translation symmetry.
Much like we have already shown for charge conjugation symmetry, translation symmetry can exhibit a tension with HSF.
In particular, while translations leave some Krylov sectors invariant, they mix others (see Fig.~\ref{fig:graph}).
Under certain conditions, this nontrivial action of translation symmetry enables the breaking of translation invariance at late times, as we now explain.

In a sufficiently large Krylov sector with ``generic" volume-law eigenstates, the dynamics from an initial product state is expected to equilibrate to the \textit{projection} of an infinite-temperature ensemble into that sector~\cite{Moudgalya21a,Yang20,Zhao25}.
In other words, one expects
\begin{align}
    \lim_{t\to\infty} \overline{\braket{Z_i(t)}} \approx Z^\infty_i\equiv\text{tr}(Z_i P_j)/\text{tr}(P_j),
\end{align}
where the overline denotes a time average and $P_j$ is the projector onto the Krylov subspace of the initial state.
If the Krylov sector is translation-invariant (and, indeed, within a fully connected symmetry sector without HSF), then the corresponding projected ensemble is also translation invariant and $Z^\infty_i$ is independent of $i$.
Thus, while a particular initial product state is not itself translation invariant, translation symmetry is restored at sufficiently late times if the dynamics equilibrates to this ensemble.
However, if the Krylov sector is not translation-invariant, then $Z^\infty_i$ can acquire a spatial dependence.
Furthermore, note that $Z^\infty_i$ is identically zero in any Krylov sector that is invariant under $X$, even if that sector is not translationally invariant.
Thus, an initial product state can equilibrate to a late time state with nontrivial spatial dependence \textit{only if} it belongs to a Krylov sector that is not closed under translation or charge conjugation.

As an example, Fig.~\ref{fig:dynamics} plots the dynamics under $U(t)$ (as in Fig.~\ref{fig:qubit-dynamics} but with $h=0$ to ensure translation symmetry) of an initial product state $\ket{\psi_0}$ on $L=24$ sites within a Krylov sector containing 1456 states.
The local expectation values $\braket{Z_i(t)}=\braket{\psi_0|U^\dagger(t)\, Z_i\, U(t)|\psi_0}$ relax to a staggered profile, consistent with the exact prediction $Z^\infty_i=(-1)^i(-1/4)$~\cite{Vuina25}.
The opposite staggering can be obtained by starting from the conjugated initial state $X\ket{\psi_0}$.
While the value of the late-time imbalance in a given Krylov sector can be easily obtained numerically, analytical calculations as in Ref.~\cite{Vuina25} require a detailed understanding of the HSF structure.
Symmetry fragmentation provides a complementary mechanistic viewpoint of this phenomenon that is agnostic of such details.

\textit{Conclusion.}---We have demonstrated that the interplay of symmetry and HSF can give rise to exponentially many emergent qubits encoded in the eigenstates of a quantum many-body system.
In the U(1)$\rtimes\mathbb Z_2$ example considered here, these qubits emerge from the pairing of certain Krylov sectors by $\mathbb Z_2$ symmetry.
Symmetry fragmentation also provides a mechanistic interpretation of the late-time density imbalances typically used to probe HSF in experiments.
It will be fascinating to explore the variety of patterns of symmetry fragmentation that can emerge.
A particularly interesting direction is the interplay with generalized, higher-form, and subsystem symmetries~\cite{Nussinov09a,Nussinov09b,Gaiotto15,McGreevy23,Khudorozhkov22,Stephen24,Stahl24,Stahl25}, which may enable still more exotic phenomena.

\begin{acknowledgments}
I am grateful to Claudio Chamon, Anushya Chandran, Jean-Yves Desaules, Paul Fendley, Sanjay Moudgalya, Rahul Nandkishore, and Justin Wilson for helpful discussions and feedback on the manuscript.
This material is based upon work supported by the National Science Foundation under Grant Number~DMR-2143635.
This work was performed in part at the Aspen Center for Physics, which is supported by National Science Foundation grant PHY-2210452, and at the Kavli Institute for Theoretical Physics, which is supported by National Science Foundation grant PHY-2309135.
\end{acknowledgments}

\bibliography{refs}




\end{document}